\documentclass[11pt]{article}
\usepackage{moriond,epsfig}
\usepackage{url}
\usepackage{color,amssymb,amsmath}

\def \Vista  {{\sc Vista}}
\def \Sleuth {{\sc Sleuth}}
\def \Quaero {{\sc Quaero}}
\def \Bard   {{\sc Bard}}
\def \TurboSim {{\sc TurboSim}}
\def \DZero {{D\O}}
\def \Aleph {{\sc Aleph}}

\def \published {{\textcolor{blue}{\ensuremath{\checkmark}}}}
\def \complete {{\textcolor{yellow}{\ensuremath{\bigstar}}}}
\def \ongoing  {{\textcolor{green}{\ensuremath{\bullet}}}}
\def \stalled  {{\textcolor{red}{{\ensuremath{\blacklozenge}}}}}

\def\be{\begin{equation}}
\def\ee{\end{equation}}
\def\bea{\begin{eqnarray}}
\def\eea{\end{eqnarray}}

\begin{document}
\vspace*{4cm}
\title{SYSTEMATIC ANALYSIS OF FRONTIER ENERGY COLLIDER DATA}

\author{ BRUCE KNUTESON }

\address{Department of Physics, Massachusetts Institute of Technology, 77 Massachusetts Avenue, \\
Cambridge 02139, USA}

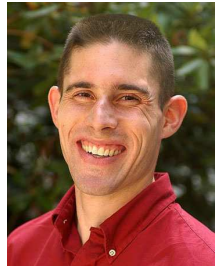
\maketitle\abstracts{
Ignorance of the form new physics will take suggests the importance of systematically analyzing all data collected at the energy frontier, with the goal of maximizing the chance for discovery both before and after the turn on of the LHC.
}

\section{The Game}

%The approach currently adopted by the frontier energy collider experiments is to choose a number of different beyond-the-standard-model hypotheses and test these against the data.  This is a fine way to proceed if the number of competing candidate theories is small, but it is a really amazingly stupid way to proceed if the number of competing candidate theories is large.

%Question to the experimentalist:  
Which of the many possible extensions to the Standard Model should you spend your time trying to detect?  The standard American technique for finding an answer to such a difficult question is to conduct a poll, since averaging over the responses of many uninformed people has been shown to produce useful insight.  In this spirit, and in consultation with Gallup, a poll has been constructed for contemporary high energy physics.  The first question of this poll is: 

\begin{center} 
\begin{tabular}{ll}
\multicolumn{2}{l}{\hspace*{-0.5in} 1. The first sign of new physics will come from (check one):} \\
$\square$  Heavy gauge bosons      &   $\square$   Technicolor    \\
$\square$  Large extra dimensions  &   $\square$   Leptoquarks    \\
$\square$ Supersymmetry            &   $\square$   a fourth generation of fermions \\
$\square$ Compositeness            &   $\square$   Something else  \\
\end{tabular}
\end{center}
The result of this poll, averaging over the responses of several hundred professional physicists, is provided in the transparencies on which these proceedings are based.\cite{KnutesonMoriondTransparencies}  As with most polls, the result of this one is not particularly informative, but it is nonetheless mildly interesting.  Although none of the options wins the majority, supersymmetry carries the plurality.  

Assuming Nature, like most Western democracies, is governed by public opinion, she has presumably chosen to be supersymmetric.  In the interest of elegance and bureaucratic efficiency, she has doubtlessly also chosen to be minimally supersymmetric.  In this restrictive case, the second question of the poll is:  

{\hspace*{0.35in} 2. What are the values of the 105 parameters of the MSSM?} \\

Your local state lottery provides a larger payout at better odds.

How will a discovery look in an Easter egg hunt involving 1000 physicists each testing one of 1000 different models?  Although each random member of each kilophysicist collaboration will guess wrong, suppose one experimentalist is very lucky and guesses close.  To understand what ``close'' means, consider a space of two observables, with a true signal indicated by some clustering of events in this space.  The physicist takes a particular model from a theoretical friend, blinds himself to the data, chooses some region in this space of observables based on that model, compares the number of events seen in that region to the number expected from Standard Model processes, and finds an excess corresponding to 3.5 standard deviations.  He then looks at the rest of the data, and realizes that if he had chosen a slightly different model, he could have found an excess corresponding to 6.5 standard deviations.  Sticky questions of interpretation naturally ensue.  Any ``successful'' dedicated search for new physics is bound to end up in a set of highly sculpted cuts, since the original guess is bound to be wrong.

The development of a systematic framework within which the entirety of the frontier energy collider data can be understood could be an important play both in the short term game of maximizing the field's chance for discovery before the turn on of the LHC, and in the medium term game of maximizing the field's chance for discovery after the turn on of the LHC.

\section{Strategy and Tactics}

The first prong of this systematic framework is an algorithm called \Vista.  \Vista\ is a model-independent search for new large cross section physics, designed for understanding the gross features of the data and the bulks of distributions.  Basic physics objects are defined:  electrons, muons, taus, photons, jets, $b$-jets, and missing energy; all high-$p_T$ events are collected; the contribution from all Standard Model processes are estimated, turning event generators into a virtual collider; the detector response is simulated; and experimental and theoretical fudge factors are systematically fit.  The data and all Standard Model backgrounds are then partitioned into exclusive final states.  The number of events in each final state is noted and compared to prediction, and the discrepancy is quantified in units of standard deviations, taking into account the several hundred different final states considered.  This list is ordered according to decreasing discrepancy, and each day's task is to understand the discrepancies at the top of the list.  Supplementing this single-page vista of the data landscape are ten thousand automatically generated plots showing a comparison between data and Standard Model prediction in all relevant kinematic distributions, with differences in shape quantified using the simple KS statistic, taking into account the several thousand different distributions considered.

The second prong of this framework is an algorithm called \Sleuth.  \Sleuth\ is a quasi-model-independent search strategy for new high-$p_T$, small cross section physics.  The bulks of distributions having been understood using \Vista, \Sleuth\ focuses on identifying excesses on the high-$p_T$ tails.  \Sleuth\ rigorously computes the infamous trials factor, quantifying the fraction of hypothetical similar experiments in which one would see something as interesting as what one actually sees in the data.  This has been achieved so far only at two of the frontier energy collider experiments.  The first was \Sleuth@\DZero RunI,\cite{SleuthPRL:Abbott:2001ke,SleuthPRD1:Abbott:2000fb,SleuthPRD2:Abbott:2000gx,KnutesonThesis} in which roughly thirty final states were analyzed in a quasi-model-independent way for new physics in Tevatron Run I.  The more recent and exhaustive H1 General Search\cite{H1GeneralSearch:Aktas:2004pz} represents the first time in the last thirty years that a frontier energy collider experiment has completely understood its high-$p_T$ data.  \Sleuth@TevatronRunII will closely resemble the H1 General Search, which significantly improves and simplifies the region-searching algorithm used in \Sleuth@\DZero RunI.

The third prong of the framework is an algorithm called \Quaero\ (Latin for ``I search for, I seek'').  \Quaero\ is an algorithm designed for a rigorous, fast, transparent, robust, and model-dependent testing of specific hypotheses.  The user interface is viewable online.\cite{QuaeroURL}  A physicist should be able to provide the events her hypothesis predicts should be produced in each of the frontier energy colliders.  \Quaero\ is designed to handle the details of testing that hypothesis against the frontier energy collider data, taking into account expert collaboration-specific knowledge.  \Quaero\ returns a single number quantifying the extent to which the data (dis)favors the hypothesis relative to the Standard Model, and figures showing how the analysis was performed.

The knowledge of each experiment's detector response is encapsulated in \TurboSim, a fast simulation that tunes itself to an experiment's full simulation.  Events run through the full simulation are used to construct a gigantic lookup table mapping the outgoing legs of Feynman diagrams to reconstructed objects in the detector.  The price of an additional $\approx10$\% systematic uncertainty buys a speedup of $\approx 10^3$ and a decoupling from each experiment's offline framework.

The fourth prong of the framework is an algorithm called \Bard, a procedure for model construction intended to automate model builders.  Starting with a particular hint from \Vista\ or \Sleuth, \Bard\ systematically generates many possible signals, introducing new particles and couplings as necessary in order to explain what is seen. \Bard\ uses \Quaero\ to determine model parameters and to quantify the goodness of fit.  The output of \Bard\ is a reasonably exhaustive list of possible new physics interpretations, ordered according to how well each one fits the data.  

The current scorecard of this project is shown in Table 1.  

\section{Implications}

These proceedings conclude with a few mildly provocative thoughts.

If CMS vigorously pursues an approach similar to that described here, while 1000 ATLAS physicists divide themselves among 1000 different models, CMS will win.

%Astronomers and astrophysicists understand their data much better than we understand ours.  Data from the flagship experiments in astrophysics and astronomy, rivaling those in HEP in complexity, are sufficiently well understood that they can be made available online.  Data from thousands of biological laboratories around the world are available for analysis online.  Your local Barnes \& Nobel contains several dozen good books introducing the burgeoning field of bioinformatics.  

An experiment's ability to make its data publicly available is an acid test of that experiment's understanding of its data.  To the extent that a collaboration really understands its data and detector response, making the data available in a generally useful form is trivial.  To the extent that a collaboration lacks a coherent, self-consistent picture of its data and detector response, making the data available for multipurpose use is nearly impossible.  

HIP has overtaken HEP.  The first two prongs of our framework (\Vista\ and \Sleuth) are designed to find discrepancies in the data worth pursuing.  Having been unsuccessful in finding such discrepancies in frontier energy collider physics for the past quarter century, the possibility of seeing such discrepancies at the LHC in a few years' time is eagerly anticipated.  Heavy ion physics is already awash in discrepancies begging to be understood.  The construction of a \Quaero-like interface to the RHIC data, and of a \Bard-like method for systematically exploring the space of possible models, represents a novel and potentially fruitful way to proceed.

By algorithmatizing the systematic identification of discrepancies, the testing of different hypotheses against the data, and the construction of new models, we are beginning to automate the scientific method in the narrow field of frontier energy collider physics.  It will be interesting to see whether this approach can be successfully generalized to other subfields of science.

\section{Summary}

This discussion rests on the realization that guessing right is impossible --- the physics seen at the TeV scale is guaranteed to be a surprise.  In light of this, a few of us are pursuing a systematic analysis of all frontier energy collider data.  \Vista\ allows a model-independent search for new large cross section physics; \Sleuth\ enables a quasi-model-independent search for new high-$p_T$, small cross section physics; \Quaero\ provides a model-dependent automation of hypothesis testing; and \Bard\ provides a procedure for model construction to aid in interpreting the underlying physics.  The point of this approach is to systematically maximize the chance for discovery, both before and after the turn on of the LHC.

\begin{table}[t]
\caption{Present project status.  The columns below list project steps; rows show experiments.  References are provided to steps that are completed and published, shown as \published.  Steps that are technically complete but not yet published are shown as \complete; ongoing work is shown as \ongoing; stalled efforts are shown as \stalled.  Steps with two displayed symbols represent efforts in Tevatron Runs I and II.  No collaboration commitments are expressed or implied.}
\vspace{0.4cm}
\begin{center}
\begin{tabular}{|cc|cccc|}
\hline
     &    &  \Vista  & \Sleuth  & \TurboSim & \Quaero \\
     &    & \multicolumn{2}{c}{(General Search)} & & \\ \hline
DESY & H1 & \published\cite{H1GeneralSearch:Aktas:2004pz} & \published\cite{H1GeneralSearch:Aktas:2004pz} & \complete & \complete \\
FNAL & D\O& \published\cite{SleuthPRL:Abbott:2001ke,SleuthPRD1:Abbott:2000fb,SleuthPRD2:Abbott:2000gx,KnutesonThesis}\stalled & \published\cite{SleuthPRL:Abbott:2001ke,SleuthPRD1:Abbott:2000fb,SleuthPRD2:Abbott:2000gx,KnutesonThesis}\stalled & \stalled & \published\cite{QuaeroPRL:Abazov:2001ny}\stalled \\
     & CDF& \ongoing & \ongoing & \ongoing & \ongoing \\
CERN & Aleph& \complete &       & \complete & \complete \\
     & L3 & \complete   &       & \complete & \complete \\ \hline
\multicolumn{2}{|c|}{Algorithms} & \complete & \published\cite{SleuthPRD1:Abbott:2000fb,H1GeneralSearch:Aktas:2004pz,ACAT2003Proceedings:Knuteson:2004nj} & \complete & \published\cite{QuaeroPRL:Abazov:2001ny}\complete \\ \hline
\end{tabular}
\end{center}
\end{table}

\section*{Acknowledgments}

Collaborators include Sascha Caron at H1; Markus Klute, Khaldoun Makhoul, and Georgios Choudalakis at CDF; Daniel Whiteson, Greg Landsberg, and Dave Toback at D\O; Kyle Cranmer at \Aleph; Andr\'e Holzner at L3; and Steve Mrenna in the construction of \Bard.  Among the members of the Moriond QCD 2005 scientific committee responsible for organizing this engaging conference, Boaz Klima and Bolek Pietrzyk have provided particularly valuable support.

The project relies on the interest and expertise of the H1 collaboration at DESY, the CDF and \DZero\ collaborations at Fermilab, and the \Aleph\ and L3 collaborations at CERN.  Computing support comes from DESY, CERN, FNAL, the University of Chicago, and the Rice Terascale Cluster.  Financial support for this effort comes in part from a Department of Defense Graduate Science and Engineering Fellowship at the University of California at Berkeley; NSF International Research Fellowship INT-0107322 at CERN; a Fermi/McCormick Fellowship at the University of Chicago; and DoE grant DE-FC02-94ER40818 at the Laboratory for Nuclear Science at MIT.  Participation in this conference was partly covered through a grant from the European Union Marie Curie Programme.

\section*{References}
%\begin{thebibliography}{99}
%\end{thebibliography}
\bibliography{moriond2005}

\end{document}